\documentclass[11pt]{article}
\usepackage{epsfig,amsmath, amssymb, cite}
\newcommand{\qq}[1]{\textquotedblleft#1\/\textquotedblright}
\newcommand{\dE}{{{\scriptstyle\Delta}E}}
\newcommand{\mpi}{m_{\pi^0}}

\begin{document}
\title{Some aspects of Dalitz decay $\pi^0 \rightarrow e^+e^-\gamma$\setcounter{footnote}{4}\thanks{Presented by K.
K. at Int. Conf. Hadron Structure '02, Herlany, Slovakia, September 22-27, 2002 }}
\author{Karol Kampf\setcounter{footnote}{1}\footnote{karol.kampf@mff.cuni.cz},
Ji\v{r}\'{\i} Novotn\'y\footnote{jiri.novotny@mff.cuni.cz}\\
Institute of Particle and Nuclear Physics, Charles University,\\V
Holesovickach 2, 180 00 Prague 8, Czech Republic \and Marc
Knecht\footnote{knecht@cpt.univ-mrs.fr} \\ Centre de Physique
Th\'eorique, CNRS-Luminy, Case 907\\F-13288 Marseille Cedex 9,
France}

\date{}\maketitle
%\vspace{-0.5cm}
\begin{abstract}A calculation of $\pi^0 \rightarrow e^+ e^- \gamma$
in next-to-leading order in chiral perturbation theory of two
flavour case extended by virtual photons is presented. The whole
kinematic sector is covered and realistic experimental situation
is discussed.
\end{abstract}
\section{Introduction}

With a branching ratio of $(1.198 \pm 0.032)\%$ \cite{pdg}
the three body decay $\pi^0 \rightarrow e^+ e^- \gamma$,
first considered by R.~H.~Dalitz more than fifty years ago \cite{dalitz51},
is the second most important decay channel of the pion.
The dominant decay mode, $\pi^0\rightarrow \gamma\gamma$, with its
overwhelming branching ratio of $\text{BR} = (98.798\pm 0.032)\%$,
is deeply connected with this three body decay.
The other decay channels connected with the anomalous $\pi^0\gamma\gamma$ vertex,
like $\pi^0 \rightarrow e^+e^-$ and $\pi^0 \rightarrow 2(e^+e^-)$, are suppressed
approximately by respective factors of $10^{-7}$ and of $10^{-5}$.
As already alluded to, the hadronic aspects of the Dalitz decay of the
neutral pion are related to the axial anomaly. Actually, the main
features of this process can be discussed within the framework of
Chiral Perturbation Theory (ChPT -- for reviews see \cite{gasser}).
In the case at hand, we need only to consider the case of two
light quark flavours, $u$ and $d$.
We shall however use the extension of ChPT to electromagnetic interaction,
i.e. including effects of virtual photons
\cite{urech}, treating e.g. the pion mass difference as
a $p^2$ effect at lowest order.

\section{Kinematics}

One may separate the contributions to the $\pi^0 \rightarrow e^+e^- \gamma$
process into two main classes. The first corresponds
to the Feynman graphs where the electron-positron pair is produced
by a single photon (Dalitz pair). The leading contribution, of
order $\alpha^3$, to the decay rate belongs to this one-photon
reducible class, which involves a semi-off-shell
$\pi^0\gamma\gamma$ vertex. The second class
of contributions corresponds to one-photon irreducible topologies.
They start with the radiative corrections to the $\pi^0
\rightarrow e^+ e^- $ process, which involves the doubly off-shell
$\pi^0\gamma\gamma$ vertex. Their contributions to the decay rate
are thus suppressed, being at least of order $\alpha^5$. For the time
being, we shall therefore not take them into account. A complete
discussion will appear elsewhere \cite{pi0gll}.
%%%%%%%%%%%%%%%%%%%%%%
%%%%%%%%%%%%%%%%%%%%%%
\begin{figure}[t]
  \begin{center}\vspace{-0.5cm}
  \scalebox{.8}{\epsfig{figure=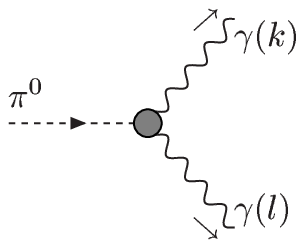}}
  \qquad \qquad
  \scalebox{.8}{\epsfig{figure=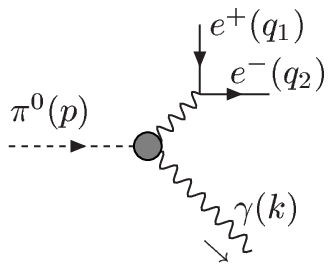}}\vspace{-0.5cm}
  \end{center}
  \caption{Kinematics lay for $\pi^0 \rightarrow \gamma\gamma$ and $\pi^0 \rightarrow e^+ e^- \gamma$}
  \label{kin}
\end{figure}
%%%%%%%%%%%%%%%%%%%%%%
%%%%%%%%%%%%%%%%%%%%%%
The general expression of the one-photon reducible part of the
$\pi^0 (p)\rightarrow e^+(q_1) e^-(q_2)$ $\gamma(k)$ amplitude is
of the form
\begin{multline}
{\cal M}_{1\gamma R}(\pi^0 \rightarrow e^+ e^- \gamma) \,=\,{\cal
M}_\nu
g^{\nu\rho}\,\frac{(-i)}{(q_1+q_2)^2}\,\frac{1}{1+{\overline{\Pi}}((q_1+q_2)^2)}\,
\\ \times{\bar u}(q_2)(ie)\Gamma_{\rho}(q_2,q_1)v(q_1) \,.
\end{multline}
In this expression, ${\cal M}_{\mu}(p,k)$ represents the semi-off-shell
$\pi^0\gamma\gamma$ amplitude, defined as
\begin{equation}
{\cal M}^\mu(p,k) \equiv
  \langle\gamma(k);{\rm out}|j^\mu(0)|\pi^0(p);{\rm in}\rangle =
  {\cal A}((p-k)^2) \varepsilon^{\nu\mu\alpha\beta}
  \epsilon_{\nu}(k)^* k^\alpha p^\beta
\,,
  \label{M1}
\end{equation}
where the second expression follows from Lorentz invariance and
parity conservation, while $j^\mu$ denotes the electromagnetic
current. Furthermore, ${\overline{\Pi}}\left( (q_1+q_2)^2\right)$ denotes the
renormalized vacuum polarization function, with
${\overline{\Pi}}(0)=0$, and
$\Gamma_{\rho}(q_2,q_1)$ stands for the vertex function,
renormalized in the on-shell scheme, of the current
$j_\rho$. Assuming again parity conservation, one may express
$\Gamma_{\rho}(q_2,q_1)$ in terms of two form factors
\begin{equation}
\Gamma_{\rho}(q_2,q_1) = F_1\left( (q_1 + q_2)^2\right)\gamma_{\rho}+
F_2\left( (q_1 + q_2)^2\right) \frac{i}{2m} \sigma_{\rho\sigma}(q_1 + q_2)^{\sigma}.
\end{equation}
The Dirac form factor $F_1$ is thus normalized by $F_1(0)=1$, while
the Pauli form factor gives the anomalous magnetic moment
of the electron, $F_2(0)=a_e$.  Finally, since $(q_1+q_2)^\mu {\cal
M}_{\mu}(p,k) = 0$, the photon propagator reduces to the
contribution of the $g^{\nu\rho}$ piece.

It is convenient to introduce the following kinematic parameters
  \begin{alignat}{2}
  x&=\frac{(q_1+q_2)^2}{\mpi^2}, \qquad &\nu^2 &\leq x \leq 1, \qquad \nu^2=\frac{4m^2}{\mpi^2}\,,\\
  y&=\frac{2p\cdot(q_1-q_2)}{\mpi^2 (1-x)},\qquad &-\beta &\leq y \leq
  \beta,\qquad \beta=\sqrt{1-\frac{\nu^2}{x}}.
  \end{alignat}
One may then write the partial decay rate, normalized to the
$\pi^0\rightarrow \gamma\gamma$ decay rate $\Gamma_0=
(1/64\pi){\cal A}(0)\mpi^3$,  for the one photon reducible
contribution in terms of these parameters as
\begin{align}
\frac{1}{\Gamma_0}\frac{d\Gamma}{{\rm d}x{\rm d}y}&=\frac \alpha
\pi \Bigl| \frac{{\cal A}(m_{\pi
^0}^2x)}{{\cal A}(0)}\frac 1{1+{\overline{\Pi }}%
(m_{\pi ^0}^2x)}\Bigr| ^2\frac{(1-x)^3}{4\,x^2}
%\quad \\ &&
\Bigl\{ \left| F_1(m_{\pi ^0}^2x)\right| ^2(\nu ^2+xy^2+x)
\notag\\ &+\left| F_2(m_{\pi ^0}^2x)\right| ^2\frac {x(\nu
^2-xy^2+x)}{\nu ^2}
%\\&&
+4{\rm Re}(F_1(m_{\pi ^0}^2x)F_2^{*}(m_{\pi ^0}^2x))x\Bigr\},
\end{align}
or, by integrating out the parameter $y$
\begin{align}
\frac{1}{\Gamma_0}\frac{d\Gamma }{{\rm d}x} &=\frac \alpha \pi
\Bigl| \frac{{\cal A}(m_{\pi
^0}^2x)}{{\cal A}(0)}\frac 1{1+{\overline{\Pi }}%
(m_{\pi ^0}^2x)}\Bigr| ^2\frac{(1-x)^3}{3\,x^2}\beta
%\quad \\ &&
\Bigl\{ \left| F_1(m_{\pi ^0}^2x)\right| ^2(\nu ^2+2x)\notag
\\&
+\left| F_2(m_{\pi ^0}^2x)\right| ^2\frac {x(2\nu ^2+x)}{\nu ^2}
%\\&&
+6{\rm Re}(F_1(m_{\pi ^0}^2x)F_2^{*}(m_{\pi ^0}^2x))x\Bigr\}.
\end{align}
At lowest order in the fine structure constant we have
${\bar\Pi}=0$, $F_1=1$ and $F_2=0$.
We shall discuss ${\cal A}$ and the hadronic part of
${\overline{\Pi}}$  in the next section.

\section{Parameterization of Hadronic Part}

We are going to study the problem in two flavour Chiral Perturbation Theory extended by
virtual photons \cite{gasser, urech, kaiser, kn} up to one loop, i.e. for this
type of process up to the ${\cal O}(p^6)$ where $p$ can represent momenta relevant for
the decay or electromagnetic constant $e$.
The lowest order effective Lagrangian can be written as
\begin{equation}
{\cal L}^{(2)}=\frac{F^2}{4}
\langle {\rm d}^\mu U^+ {\rm d}_\mu U + \chi^+ U +U^+ \chi \rangle
-\frac{1}{4} F^{\mu\nu}F_{\mu\nu}
+ Z F^4 \langle Q_R U Q_L U^+ \rangle
\end{equation}
with
\begin{equation}
U= {\rm e}^{\frac{i\phi}{F_0}}, \quad \phi= \left(
  \begin{array}{cc} \pi^0 & \sqrt{2}\pi^+\\ \sqrt{2}\pi^- & -\pi^0 \end{array} \right).
\end{equation}
Further chiral invariant Lagrangians of higher order
(${\cal O}(p^4)$ and ${\cal O}(p^6)$), so-called Wess-Zumino anomaly Lagrangian
(of ${\cal O}(p^4)$ order) and also the other details of calculation within ChPT
can be found in above references.

%\newsavebox{\vertex}
% \savebox{\vertex}{\scalebox{0.7}{\epsfig{file=vertex.eps}}}

\begin{figure}
  \begin{center}\vspace{-0.5cm}
  \epsfig{figure=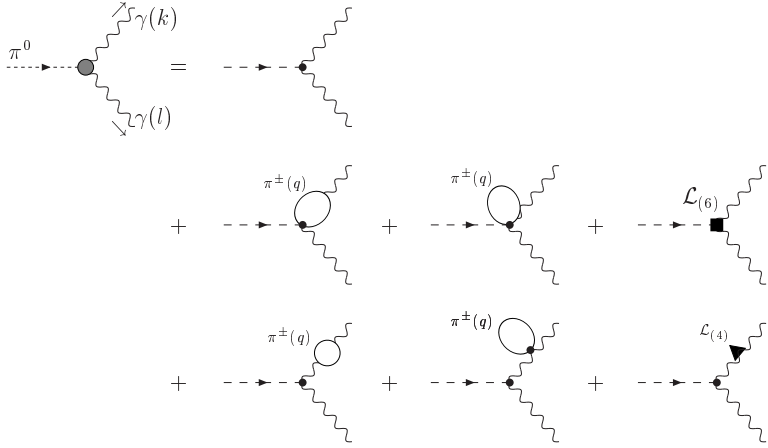, width=0.8\textwidth}\vspace{-0.5cm}
  \end{center}
  \caption{{\small Leading order and pion loop contributions to the amplitude
  ${\cal M}_{1\gamma R}\propto {\cal
A}(1-\bar{\Pi}_{had}) $. To simplify the figure, we have omitted
the $e^+e^-\gamma$-vertex. The graphs in the second row correspond
to the corrections to ${\cal A}$, while the third row contributes
by the factor $1-\bar{\Pi}_{had}$.}}
  \label{vertex}
  \end{figure}

We will calculate the hadronic part of one photon reducible
amplitude of the decay $\pi^0 \rightarrow e^+e^-\gamma$, as it is
depicted in Fig.\ref{vertex}, {\em i.e.} the value of ${\cal
A}(1-\bar{\Pi}_{had})$ in the terms of the previous section.
With definition $\sigma\equiv \sqrt{1-4m_{\pi^\pm}^2/{l^2}}$ one
can obtain from the one photon irreducible graphs in
Fig.\ref{vertex}
\begin{multline}\label{Mo}
{\cal A}(l^2) = -\frac{N_C e^2}{12\pi^2F_{\pi^0}}
 \Bigl\{
1-\frac{l^2}{6F_0^2}\frac{1}{(4\pi)^2} \Bigl[\log
\frac{m_{\pi^\pm}^2}{m_{\pi^0}^2} + \tfrac{1}{3}-2\sigma^2
\\-\sigma^3
\log\frac{\sigma-1}{\sigma+1}-\overline{{\scriptstyle\Delta}a}
\Bigr] + \delta a\Bigr\},
\end{multline}
while the one particle reducible graphs contribute by the factor
$(1-\bar{\Pi}_{had})$, where the hadronic contribution to
renormalized vacuum polarization function is
\begin{equation}
\label{pihad}
\bar{\Pi}_{had}(l^2)=\frac{e^2}{48\pi^2}\Bigl[\tfrac{2}{3}+2\sigma^2
+\sigma^3 \log\frac{\sigma-1}{\sigma+1} \Bigl]\Bigr\}.
\end{equation}
Here we have used standard handling with infinities ($\lambda$)
for ${\cal O}(p^6)$ constants:
\begin{align}
{\scriptstyle\Delta}a &= {\scriptstyle\Delta}a^r + {\scriptstyle\Delta}\alpha \lambda \notag\\
\delta a &= \delta a^r
\end{align}
with renormalized quantities to be finite and
\begin{equation}\label{da1}
{\scriptstyle\Delta}a^r = \frac{{\scriptstyle\Delta}\alpha}{32\pi^2} \bigl( \overline{{\scriptstyle\Delta}a} + \log\frac{\mpi^2}{\mu^2} \bigr),
\qquad
{\scriptstyle\Delta}\alpha = - \frac{1}{12\pi^2}.
\end{equation}
The ${\cal O}(p^6)$ constants used in this text could be connected with
form of \cite{fearing} and \cite{bijnens} respectively as
\begin{alignat}{2}\label{da2}
{\scriptstyle\Delta}a &= \frac{8}{3}F_0^2 (A_2 - 4 A_3) = \frac{8}{3}F_0^2 c_{13} \\
\delta a &= \frac{32}{3} \pi^2 \Bigl[(A_2 - 2 A_3 -4 A_4)m_{\pi^0}^2
 + \frac{20}{3}(A_4+2 A_6) 2\bar{m}B\Bigr]=\notag\\
 &= \frac{64}{3} \pi^2 \Bigl[ (c_{11} - 4c_3 -4 c_7)m_{\pi^0}^2 +
 \frac{4}{3}(5c_3 + c_7 +2 c_8)2\bar{m}B\Bigr].
\end{alignat}

The influence of virtual photon is covered in (\ref{Mo})
in the neutral pion decay constant $F_{\pi^0}$ \cite{urech}.
Unfortunately this constant is not known very accurately
$F_{\pi^0} = 92\pm 4$ \cite{pdg} and so in practice the charged pion decay constant
is used $F_\pi = 92.4$
\begin{equation}
F_{\pi^0} = F_\pi \Bigl(%
1 -\frac{m_{\pi^\pm}^2}{16 \pi^2 F_0^2} \log\frac{m_{\pi^\pm}^2}{m_{\pi^0}^2}
-\frac{e^2}{64\pi^2}\bigl[
(3+\tfrac{4}{9}Z)\bar{k}_1 - \tfrac{40}{9} Z \bar{k}_2 -3 \bar{k}_3 -4 Z
\bar{k}_4 \bigr] \Bigr).
\end{equation}

To cover this section we will introduce a {\it slope\/} parameter
$a$ which is a parameter of linear expansion of ${\cal
A}(1-\bar{\Pi}_{had})$ in $x$:
\begin{equation}\label{Ma}
{\cal A}(1-\bar{\Pi}_{had}) \sim (1 + a x)
\end{equation}
\section{Calculation at $e^3(1+p^2+e^2)$ order}

Now, to the above {\it hadronic\/} contribution we have to add also the
QED corrections, which are in our case up to one loop level:
lepton part of vacuum polarization function, vertex,
bremsstrahlung and 2-photon triangle and box corrections. In the
first approximation we will neglect 2-photon loop diagrams (for a
polemic whether this is justified see also \cite{tupper}). For
one-photon exchange the following interesting relation holds
(c.f.\cite{lautrup})
  \begin{equation}
  \Gamma (\pi^0 \rightarrow e^+e^-\gamma)=2\int_0^1 \frac{dx}{x}\frac{1}{\pi}{\rm Im}\Pi(s)\Gamma_0(s).
  \label{wbdr}
  \end{equation}
with the lepton part of spectral function of the photon
propagator~$\Pi(s)$. With ${\cal O}(p^6)$ constants set equal zero
and therefore (from (\ref{Mo}), (\ref{pihad}) and (\ref{Ma}))
$a=-0.0024$ we can obtain numerically
\begin{equation}\label{tdr}
\frac{\Gamma(\pi^0 \rightarrow e^+e^-\gamma)}{\Gamma(\pi^0\rightarrow\gamma\gamma)}
= 0.01185 + 0.00010 = 0.01196,
\end{equation}
where the second number represents loop-correction contributions.
Comparing this with PDG's number
$\Gamma^{exp}(\pi\rightarrow e^+e^-\gamma)/\Gamma(\pi\rightarrow\gamma\gamma)=(1.213\pm0.033)\%$
one can immediately see that the total decay rate is not very suitable
for probing the smooth effects of the higher orders
(e.g. tuning ChPT constants).

Therefore we will turn our attention to the differential decay rate,
which can be calculated in one-photon case using the relation (obtained from (\ref{wbdr}))
\begin{equation}\label{dGb}
\frac{{\rm d}\Gamma^{\rm rad}}{{\rm d}x} = \frac{2}{x}\Pi_a(x)\Gamma_0(x) +
  \int_{(\sqrt{x}+m_\gamma^{IR}/\mpi)^2}^1 \frac{{\rm d}E}{E} \Pi_b(E,x)\Gamma_0 (E),
\end{equation}
where we have divided the spectral function to the virtual and real part
\begin{equation}
  \frac{1}{\pi}{\rm Im}\Pi^{(4)}(p) = \Pi_a(p^2) + \Pi_b(p^2).
\end{equation}

\begin{figure}[h]
\begin{flushleft}
\vspace{-0.5cm}
  \epsfig{figure=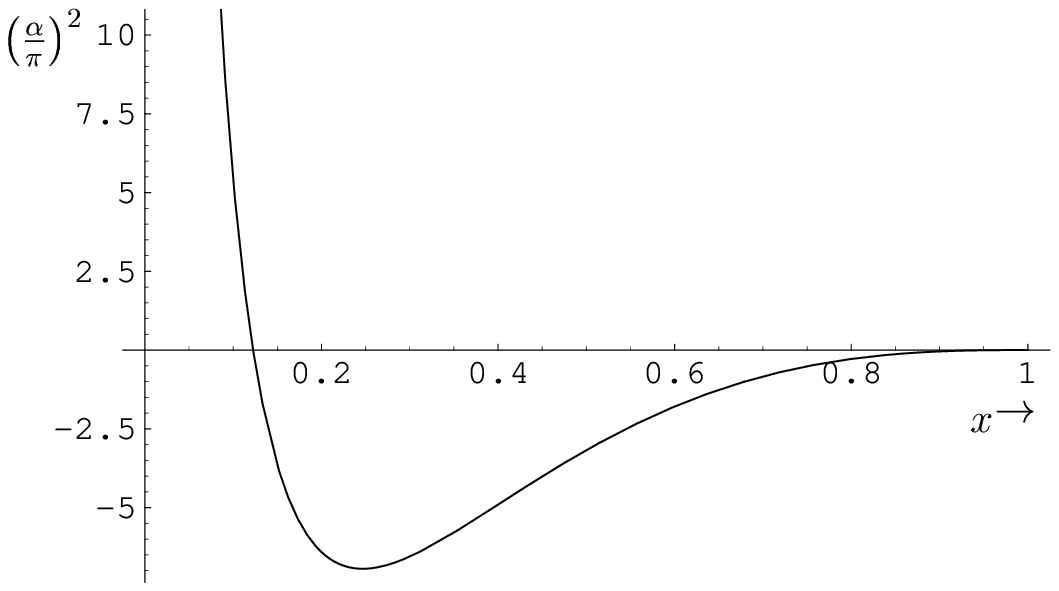,width=0.49\textwidth}
  \epsfig{figure=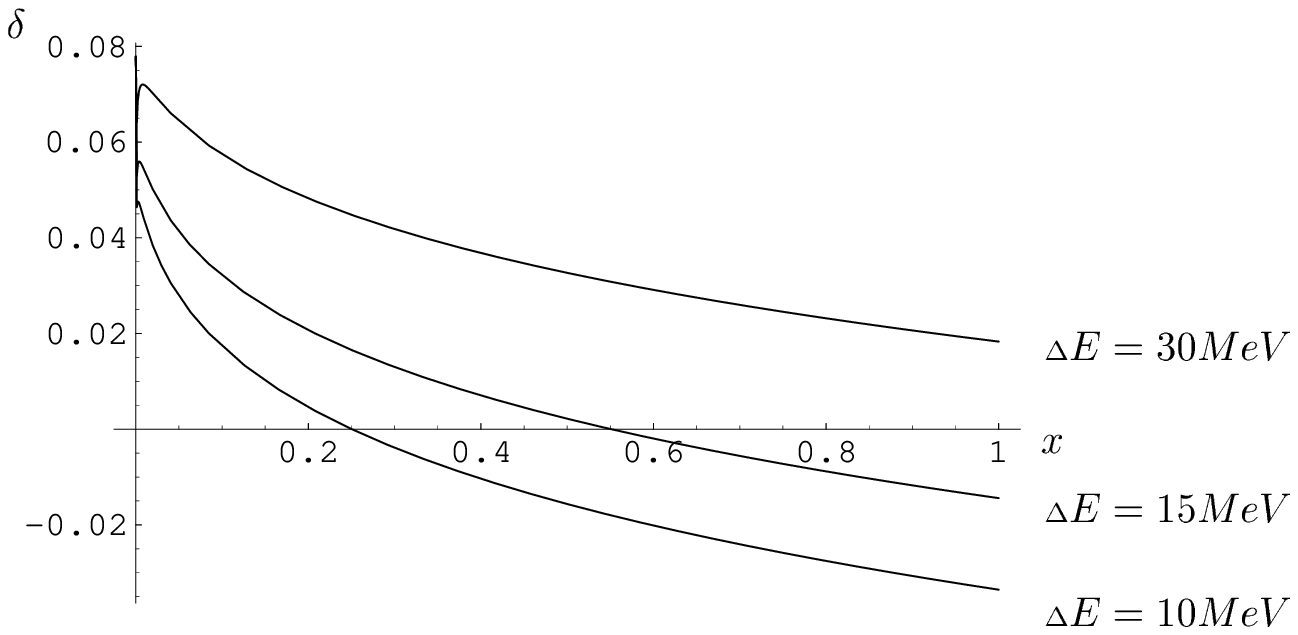,width=0.49\textwidth}
\vspace{-0.5cm}
\end{flushleft}
  \caption{Differential decay rate $\frac{1}{\Gamma_0}\frac{{\rm d}\Gamma^{(rad)}}{{\rm d}x}$ (from (\ref{dGb}))
   and the dependence of $\delta(x)$ on $\dE$}
  \label{graph1}
  \end{figure}

From the first plot in Fig.\ref{graph1} we see that while the
radiative effects for the total decay rate is small, the situation
in the differential decay rate is opposite.

However, this approach is useful for a situation when the detector is entirely blind to photons.
In a more realistic situation, the experimental resolution is such that
{\it it is possible\/} to detect photons if only above the (small) energy $\dE$.
For this case we define $\delta(x,y)$ and $\delta(x)$
\begin{equation}
  \frac{1}{\Gamma_0}\frac{{\rm d}^2\Gamma^{\rm rad}}{{\rm d}x{\rm d}y}=
  \delta(x,y)\frac{1}{\Gamma_0}\frac{{\rm d}^2\Gamma}{{\rm d}x{\rm d}y},
  \qquad
  \frac{1}{\Gamma_0}\frac{{\rm d}\Gamma^{\rm rad}}{{\rm d}x}=
  \delta(x)\frac{1}{\Gamma_0}\frac{{\rm d}\Gamma}{{\rm d}x}.
  \label{dx}
\end{equation}
In Fig.\ref{graph1} (the second plot) we depicted how the
$\delta(x)$ changes with parameter $\dE$.
 %\begin{figure}[td]
 % \begin{center}
 % \epsfig{figure=graph3.eps,width=0.4\textwidth}
 % \end{center}
 % \caption{The dependence of $\delta(x)$ on $\dE$}
 % \label{graph3}
 % \end{figure}

Up to now we have systematically neglected the graphs which arise
from the two-virtual-photons exchange. We have performed this
calculation and we found interesting behaviour of contribution to
$\delta(x)$, particularly when $x$ is close to 1. In this domain
\begin{equation}\label{deltax}
\delta(x\sim 1,y) \approx 0.1,
\end{equation}
i.e. it represents approximately 10\% of leading order and from
the previous we could conclude that it is a important contribution
to differential decay rate, which cannot be neglected. However, it
will be difficult to test this area experimentally. Actually for
\begin{equation}
x \gtrsim 1-\frac{2\dE}{m_{\pi^0}}
\end{equation}
the photon is effectively immeasurable by definition and precisely this
is the domain where the interesting situation of (\ref{deltax}) occurs.
Let us also stress that in this part of calculation we have neglected the electron mass
and thus, at this level, it is not possible to compare and use in our consideration process
$\pi^0\rightarrow e^+ e^-$ which is proportional to $m_e$ (for more details see \cite{knecht}).

\section{Conclusion}
We have discussed the Dalitz decay $\pi^0\rightarrow e^+e^-\gamma$
in two flavour ChPT and set the hadronic part within this
effective theory. Using PDG's value for the slope parameter (see
(\ref{Ma})) $a=0.032\pm 0.004$ \cite{pdg} one can estimate
following linear combination of ${\cal O}(p^6)$ constants
$$ \frac{m_{\pi^0}^2}{6
F_0^2}\frac{1}{(4\pi)^2}\overline{{\scriptstyle\Delta}a}-a \delta
a = a+0.0024
$$
(cf. (\ref{da1}) and (\ref{da2})).
We approve that the effects of higher orders in total decay rate are smaller than
the present experimental error (\ref{tdr}).
However, differential decay rate is much sensitive to these effects.\\
{\bf Acknowledgement.\ } This work was supported in part by the program
\qq{Research Centres} project number LN00A006 of the Ministry of
Education of the Czech Republic, Socrates/Erasmus programme and
EC contract HPRN-CT-2002-00311 (EURIDICE).

\end{document}